\documentclass[aps,prb,twocolumn,reprint,superscriptaddress]{revtex4-1}
\usepackage{pdfpages}
\usepackage{pgffor}
\usepackage{textcomp} % for \textcelsius{}
\usepackage{graphicx}
\usepackage[utf8]{inputenc}
\usepackage{url}
\usepackage{float}
\usepackage{hyperref}
\hypersetup{
    colorlinks=true,
    urlcolor=blue,
    linkcolor=black,
    citecolor=black
    }
    
\makeatletter
\AtBeginDocument{\let\LS@rot\@undefined}
\makeatother

\begin{document}

\author{C. M. Polley}
\email{craig.polley@maxiv.lu.se}
\affiliation{MAX IV Laboratory, Lund University, Fotongatan 2, 22484 Lund, Sweden}
\affiliation{Department of Microtechnology and Nanoscience, Chalmers University of Technology, 412 96, Gothenburg, Sweden}
\author{H. Fedderwitz}
\author{T. Balasubramanian}
\author{A. A. Zakharov}
\affiliation{MAX IV Laboratory, Lund University, Fotongatan 2, 22484 Lund, Sweden}
\author{R. Yakimova}
\affiliation{Department of Physics, Chemistry and Biology (IFM), Linköping University, Linköping, 581 83 Sweden}
\author{O. Bäcke}
\author{J. Ekman}
\author{S. P. Dash}
\author{S. Kubatkin}
\author{S. Lara-Avila}
\affiliation{Department of Microtechnology and Nanoscience, Chalmers University of Technology, 412 96, Gothenburg, Sweden}

\title{Bottom-up growth of monolayer honeycomb SiC}

\begin{abstract}

The long theorized two-dimensional allotrope of SiC has remained elusive amid the exploration of graphenelike honeycomb structured monolayers. It is anticipated to possess a large direct band gap (2.5~eV), ambient stability, and chemical versatility. While $sp^{2}$ bonding between silicon and carbon is energetically favorable, only disordered nanoflakes have been reported to date. Here we demonstrate large-area, bottom-up synthesis of monocrystalline, epitaxial monolayer honeycomb SiC atop ultrathin transition metal carbide films on SiC substrates. We find the 2D phase of SiC to be almost planar and stable at high temperatures, up to 1200\textcelsius{} in vacuum. Interactions between the 2D-SiC and the transition metal carbide surface result in a Dirac-like feature in the electronic band structure, which in the case of a TaC substrate is strongly spin-split. Our findings represent the first step towards routine and tailored synthesis of 2D-SiC monolayers, and this novel heteroepitaxial system may find diverse applications ranging from photovoltaics to topological superconductivity.
\end{abstract}


\maketitle
Chemical bonding between C and Si may exhibit competition between the $sp^{3}$ hybridization preferred by silicon and the $sp^{2}$ hybridization preferred by carbon\cite{Melion2007}. Nonetheless, first-principles calculations predict that two-dimensional, $sp^{2}$ bonded SiC (2D-SiC) should be stable, with the planar, stoichiometric structure the most energetically favorable conformation\cite{Sahin2009,Huda2009,Zhou2013,Shi2015}. Despite considerable interest in its properties and several theoretical studies\cite{Chabi2020,Lu2012,Lin2013,Hsueh2011}, experimental synthesis of 2D-SiC has proven difficult\cite{Lin2012,Susi2017,Chabi2021}. Central to this challenge is that, unlike graphitic carbon, bulk SiC does not exist in a layered form and hence does not lend itself to top-down exfoliation techniques.

Here we demonstrate a bottom-up approach to the synthesis of 2D-SiC, by preparing ultra thin carbide films (t$<$3~nm) on SiC substrates followed by annealing to high temperature (T$>$1700\textcelsius{}), similar to the process used for epitaxial growth of graphene on SiC. Underlying this approach is our ability to produce high-quality ultra thin epitaxial films of transition metal carbides on the silicon terminated face of 4H-SiC. Carbides are well known for their extreme thermal, chemical and mechanical robustness, and among the transition metal carbides TaC and NbC in particular have received renewed attention owing to their bulk superconductivity (T$_{C}$ $\approx$11~K) and topologically nontrivial electronic band structure \cite{Shang2020,Cui2020,Yan2021}. Both TaC and NbC can grow epitaxially on 4H-SiC, as they have a rocksalt crystal structure with a threefold symmetric (111) plane that is nearly lattice matched to 4H-SiC (0001) (2.3\% and 2.7\% lattice mismatch, respectively). Huder $\textit{et. al.}$ reported a bottom-up solid state reaction in which films of Ta on a SiC substrate were transformed into graphene-terminated TaC by thermal annealing, with carbon being sourced from the decomposing SiC \cite{LeQuang2017,Huder2018}. While the simplicity of this approach is appealing, the resulting TaC film is polycrystalline with grain sizes of $\approx$200~nm$^{2}$. We have developed an alternative process involving sputter deposition of approximately ten monolayers of transition metal carbide followed by annealing to 1700\textcelsius{} in an Ar atmosphere. Directly depositing a carbide rather than the base metal ensures a stoichiometric film and recrystallization upon annealing. Remarkably, we observe that while the annealing step indeed structurally orders the carbide, in addition it leaves the surface terminated with a monolayer of honeycomb SiC.

\begin{figure}
    \centering
	\includegraphics[width=8.6cm]{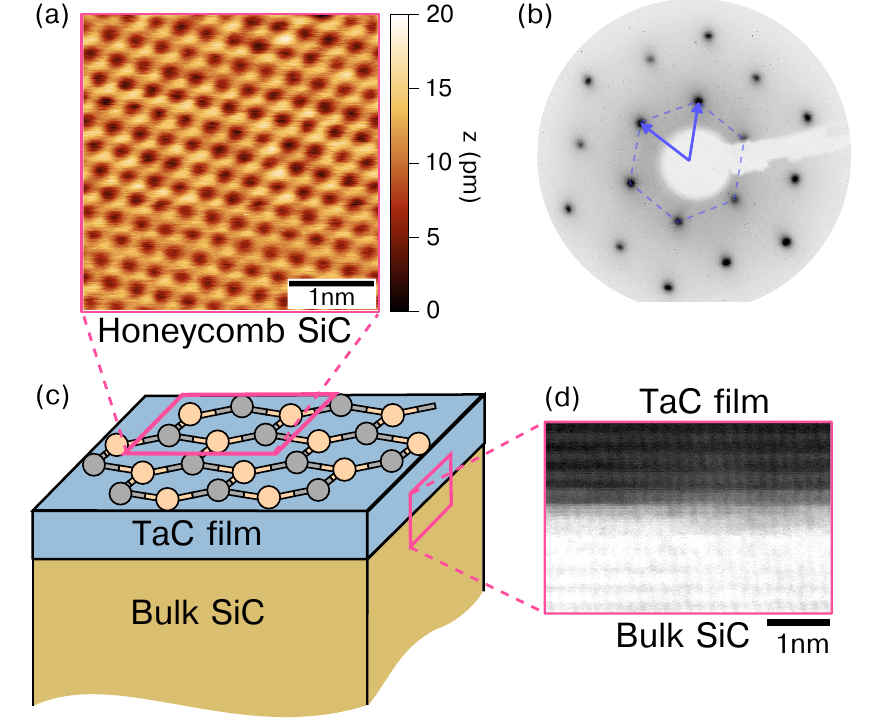}
	\caption{
		\footnotesize{(a) Topographic STM image of the monolayer honeycomb SiC layer terminating a TaC(111) surface (V$_\textrm{bias}$ = -1.2~V,  {I} = 3~nA). (b) Representative LEED image (E=145~eV); 2.32\AA$^{-1}$ unit cell reciprocal vectors are indicated. (c) Schematic depiction of the sample structure (d) Cross-sectional HAADF-STEM image of the TaC/SiC interface.
		}
	}
	\label{fig:preparation}
\end{figure}
\section*{Results}
\begin{figure}	
    \centering
	\includegraphics[width=8.6cm]{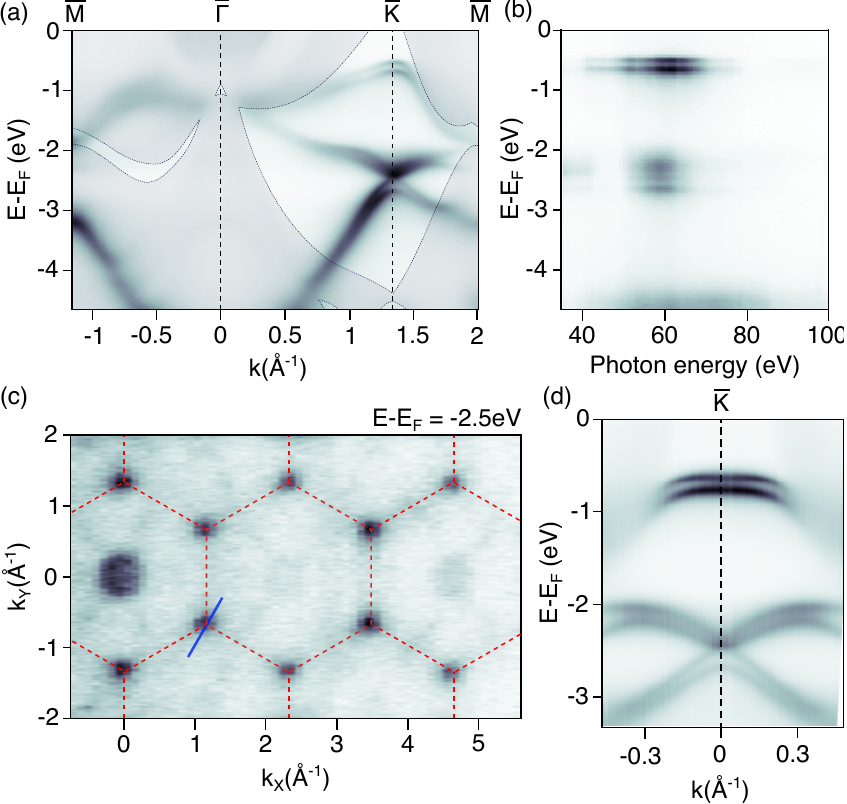}
	\caption{
		\footnotesize{(a) ARPES measurements (h$\nu$=67~eV, T=80~K) along the $\bar{M}$-$\bar{\Gamma}$-$\bar{K}$-$\bar{M}$ path, overlaid with the surface-projected bulk band structure of TaC. (b) h$\nu$ resolved cut at $\bar{K}$ (h$\nu$=35-100eV, T=80K). (c) Constant-energy surface through the Dirac-like crossing of the surface state, with the surface Brillouin zone overlaid (h$\nu$=180~eV, T=80~K). (d) High-resolution ARPES measurement of the surface states at $\bar{K}$, direction indicated by the blue line in (c) (h$\nu$=20~eV, T=19~K).}
	}
	\label{fig:ARPES1}
\end{figure}

The samples studied are schematically depicted in Figure \ref{fig:preparation}a. They consist of 3~nm thick films of TaC or NbC sputter deposited onto Si-terminated 4H-SiC(0001) substrates, with a lateral extent of approximately 5~mm $\times$ 5~mm. The deposition was followed by a 1700\textcelsius{} anneal for 10 min in an Ar atmosphere. Here we mainly focus on TaC, but later will show that a similar result is obtained from NbC. The TaC/SiC interface is abrupt and the films are epitaxial, as revealed by cross-sectional high-angle annular dark-field scanning transmission electron microscopy (HAADF-STEM). Cubic TaC has a rocksalt crystal structure with lattice parameter a=4.456\AA\cite{Toth1971}, which presents \{111\} facets with an in-plane lattice parameter a=3.15\AA, readily accommodated atop the SiC substrate with a=3.08\AA. \cite{Stockmeier2009}. While hexagonal (Ta$_2$C) or ternary (Ta$_2$SiC) phases are in principle possible, there are several indications that the films as grown take the simple rocksalt structure. The HAADF-STEM (Figure \ref{fig:preparation}d) shows cubic ordering, and the measured superconducting transition temperature of 9-10~K (see Supplementary Material) is consistent with cubic, stoichiometric TaC \cite{Giorgi1962}. After transferring the samples into UHV and briefly annealing at 1000\textcelsius{} to remove surface oxidation, sharp hexagonal spots in LEED (Figure \ref{fig:preparation}b) and bands in ARPES (Figure \ref{fig:ARPES1}) become visible. Scanning tunneling microscopy (STM) (Figure \ref{fig:preparation}a) indicates a hexagonal surface lattice with a=(3.12$\pm$0.04)\AA, uniformly covering the millimeter-scale sample (see Supplementary Material for additional topographic images) . This surface is robust against longer or repeated heating cycles to similar temperatures, but at much higher temperatures will begin to transform into graphene (see Supplementary Material).

\begin{figure*}
    \centering
	\includegraphics[width=17.2cm]{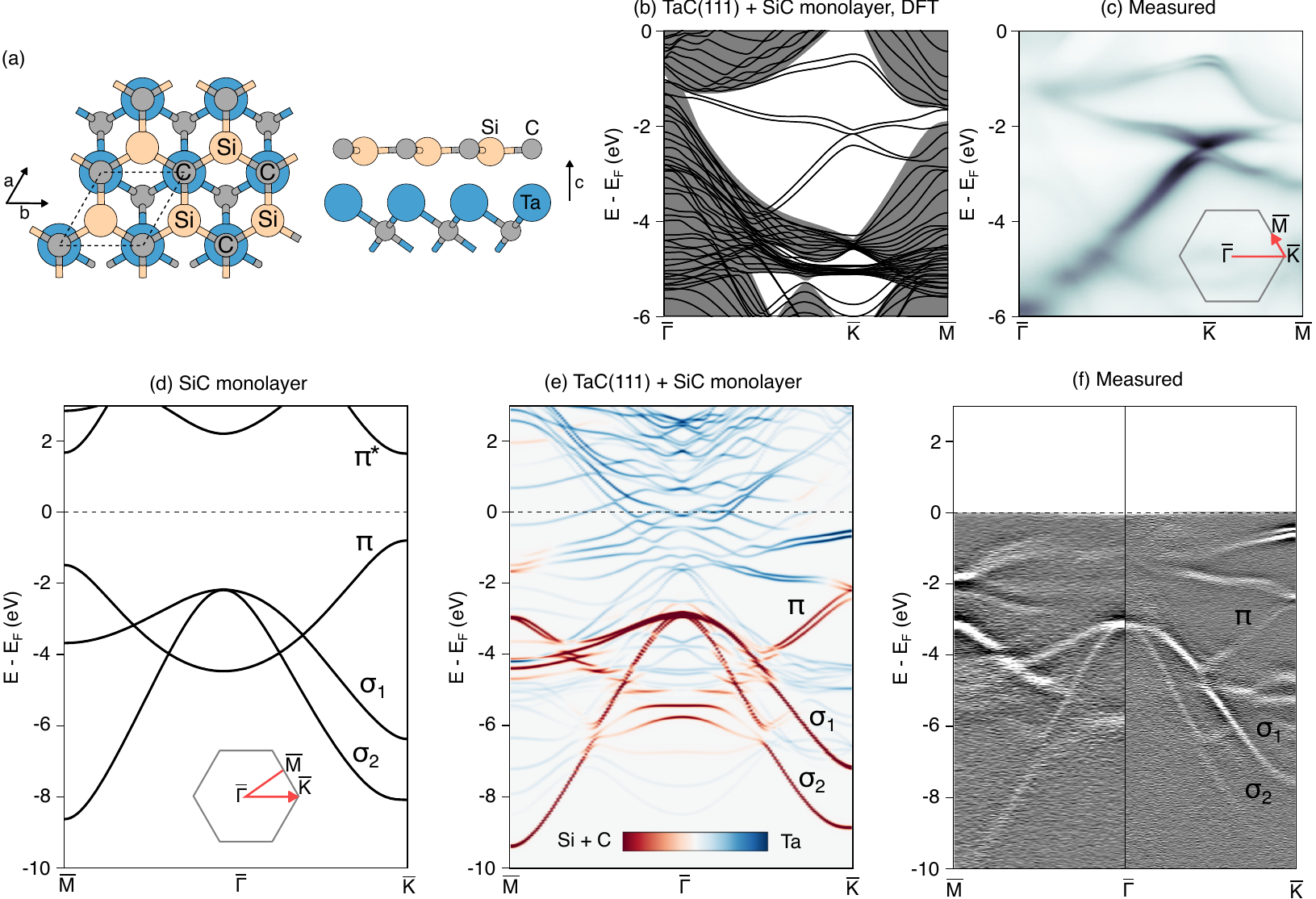}
	\caption{
		\footnotesize{(a) Schematic depiction of the honeycomb SiC structure. (b) Slab DFT band structure calculation for this structure along the $\bar{\Gamma}$-$\bar{K}$-$\bar{M}$ path. The surface-projected TaC bulk band structure is indicated by the shaded grey regions. (c) ARPES spectrum at h$\nu$=67~eV, T=80~K. (d) DFT bandstructure of a planar, freestanding SiC monolayer along the $\bar{M}$-$\bar{\Gamma}$-$\bar{K}$ path. (e) The same calculation for SiC atop TaC(111) as shown in (b), but sampling only the topmost 2 atomic layers and resolved into atomic character. (f) Corresponding experimental spectra, presented as a second derivative image. The $\bar{\Gamma}$-$\bar{M}$ slice is acquired from the second Brillouin zone with h$\nu$=138~eV while the $\bar{\Gamma}$-$\bar{K}$ slice is with h$\nu$=44~eV, both at T=80~K}
	}
	\label{fig:DFT}
\end{figure*}
The most striking features in the ARPES measurements are spin-split surface states within a large surface-projected band gap of the TaC (Figure \ref{fig:ARPES1}a). Photon energy scans (Figure \ref{fig:ARPES1}b) demonstrate the 2D character of these states, while constant-energy surfaces (Figure \ref{fig:ARPES1}c) confirm a 1$\times$1 hexagonal surface lattice (overlaid in red). The zone size ($\bar{\Gamma}$-$\bar{K}$ = 1.34\AA$^{-1}$) agrees with the lattice constant observed in STM. A high-resolution cut through $\bar{K}$ (Figure \ref{fig:ARPES1}d) clearly illustrates the spin-orbit splitting of the bands, which reaches 125~meV for the upper bands and 220~meV for the lower set. All of these states are completely occupied, with the Fermi surface dominated by TaC bulk states spanning most of the surface Brillouin zone. 

Based on evidence offered by LEED, STM, core levels (described below) and particularly DFT band structure calculations, we conclude that the ARPES measurements can only be explained by the presence of a monolayer of honeycomb SiC. The proposed structure is depicted in Figure \ref{fig:DFT}a, and generates a calculated band structure that is in excellent agreement with experimental spectra (Figure \ref{fig:DFT}b,c). The registry of the SiC monolayer has carbon atoms atop the substrate Ta atoms and silicon atoms in an fcc-hollow position. After relaxation along the c axis the SiC layer is weakly buckled ($\Delta z$ = 16.8~pm) and located 2.3\AA{} above the surface layer of Ta atoms. In order for the DFT model to reproduce the degeneracy at $\bar{K}$ the buckling must be reduced to 11.8~pm (see Supplementary Information). Unlike silicene (44~pm), the 2D-SiC here is nearly planar.

To interpret the band structure calculations it is helpful to draw comparisons with an isolated, planar honeycomb SiC monolayer (Figure \ref{fig:DFT}d). Quite unlike its semimetallic relatives graphene or (planar) silicene, honeycomb SiC features a large band gap between the $\pi$ and $\pi$* bands. In Figure \ref{fig:DFT}e we plot the atom-, layer- and $k$-resolved band structure, obtained by projecting the calculated wave functions onto the topmost two atomic layers of the slab (Si, C, and Ta atoms). This suppresses the bulklike TaC slab bands and thereby mimics the surface sensitivity of an ARPES measurement. We additionally use a color scale to represent the atomic character of each state. Presented in this manner, the SiC $\sigma _1$ and $\sigma _2$ bands become obvious both in the calculations and in experimental spectra (Figure \ref{fig:DFT}g). We can furthermore identify the valence band of the Dirac-like surface state as deriving from the SiC $\pi$ band, while the conduction band and also the uppermost surface state have strong Ta 5d orbital character. The hybridization of the $\pi$ band, together with the large spin-orbit splitting, makes it clear that the SiC layer is strongly interacting with the TaC substrate.

\begin{figure}	
    \centering
	\includegraphics[width=8.6cm]{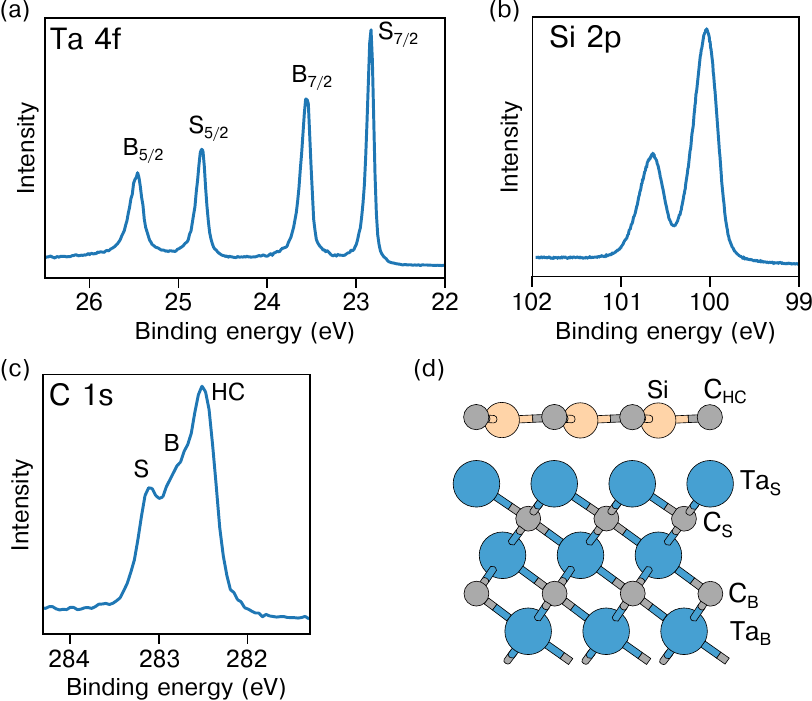}
	\caption{
		\footnotesize{High-resolution core level spectra. (a) Ta 4f at h$\nu$=140~eV ($\Delta$E=10~meV), (b) Si 2p at h$\nu$=130~eV ($\Delta$E=5~meV) and (c) C 1s at h$\nu$=430~eV ($\Delta$E=80~meV). All spectra acquired at T=20~K. (d) Suggested origin of the observed components.}
	}
	\label{fig:XPS}
\end{figure}

Core level spectroscopy measurements (Figure \ref{fig:XPS}a-c) are consistent with the proposed structure in Figure \ref{fig:DFT}a. The only species observed on the surface are Ta, Si and C, and none are trace impurities - when measured together at h$\nu$ = 650~eV the cross-section adjusted intensity ratio is Ta:Si:C = 0.4:0.25:1. The Ta 4f core level spectrum (Figure \ref{fig:XPS}a) consists of bulk ($B$) and surface ($S$) doublets, consistent with previous studies of \{111\} facets of bulk transition metal carbides that uniformly find a metal element rather than carbon termination \cite{Johansson1995}. While the position of our $B$ doublet is in excellent agreement with stoichiometric bulk TaC \cite{Gruzalski1986}, the bulk-surface doublet splitting is approximately 250~meV smaller than would be expected from a simple TaC(111) surface \cite{Johansson1995}. The Si 2p spectrum (Figure \ref{fig:XPS}b) consists of a single doublet and derives from silicon atoms close to the surface. This photon energy (130~eV) could in principle be weakly sensitive to silicon at the buried interface with the bulk SiC substrate, but this is unlikely to be the origin of the Si peak given that it shows an abrupt shift in binding energy when the surface oxide is removed (Supplementary Material). The C 1s spectrum (Figure \ref{fig:XPS}c) contains three components, of which the component $HC$ alone is closely correlated with the presence of the honeycomb SiC band structure. The other two components exist both before the SiC termination is established and after it is eliminated (see Supplementary Material). One of these peaks ($B$) can be assigned to bulk carbon in the TaC film, with a binding energy in good agreement with measurements from the (001) surface of a bulk TaC crystal\cite{Gruzalski1986}. Figure \ref{fig:XPS}d suggests origins for each component. 

\section*{Discussion}
\begin{figure}	
    \centering
	\includegraphics[width=8.6cm]{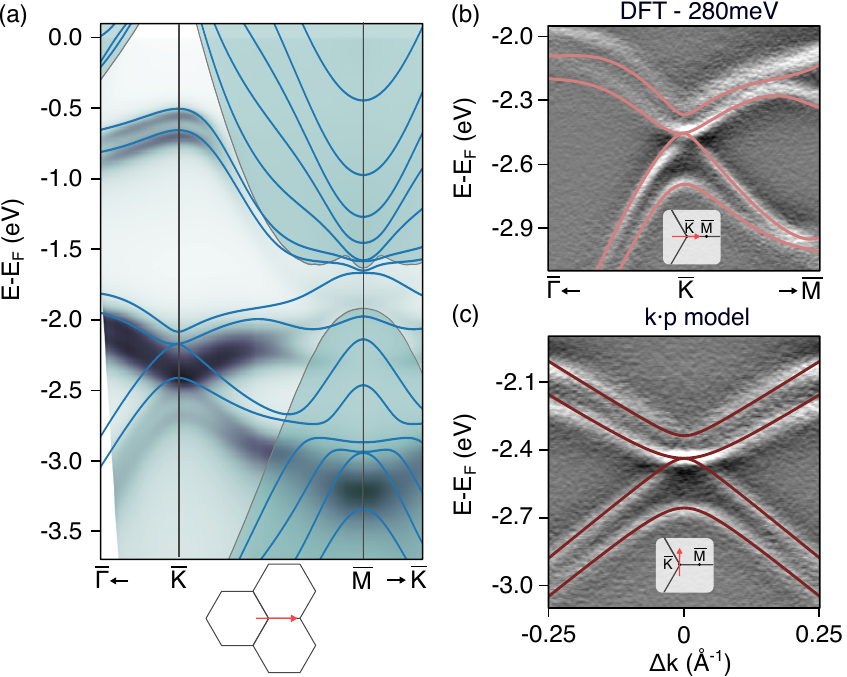}
	\caption{
		\footnotesize{(a) ARPES spectrum (h$\nu$=23~eV) overlaid with calculated band structure; shaded regions indicate surface-projected bulk bands of the TaC film. (b) Close up of the Dirac-like crossing at $\bar{K}$ (h$\nu$=40~eV), presented as a second derivative image and overlaid with the DFT band structure after applying an energy offset of 280~meV. (c) Further magnified spectrum of the Dirac-like crossing (h$\nu$=20~eV, from Fig\ref{fig:ARPES1}d), now overlaid with a fit to a k$\cdot$p model for graphene with spin-orbit coupling. All spectra acquired at T=20~K with $\Delta$E=5~meV }
	}
	\label{fig:ARPES2}
\end{figure}

While the overall agreement between the DFT calculations and ARPES measurements is excellent, there are two notable exceptions concerning the Dirac-like feature at $\bar{K}$. The first concerns the experimentally observed degeneracy - after full relaxation of the slab along the c-axis, band calculations rather predict a sizeable band gap of 200~meV. This can be resolved by reducing the buckling of the SiC monolayer from 16.8~pm to 11.8~pm, which is still larger than the $<$5~pm suggested by STM (see Supplementary Material). While it is interesting that the real system apparently assumes precisely the buckling that results in a band degeneracy at $\bar{K}$, the gap being a continuously tunable function of the buckling argues against the existence of a symmetry protection. The second exception concerns the binding energy of this feature, which requires a rigid shift of 280~meV to higher values to agree with experiments (Figure \ref{fig:ARPES2}c). The Ta 5d derived surface state at lower binding energy does not require such a shift. We assume that these discrepancies can be resolved with a more sophisticated calculation method (e.g. DFT+U) or with a more complex structural model (e.g. including a buffer layer).

\begin{figure}	
    \centering
	\includegraphics[width=8.6cm]{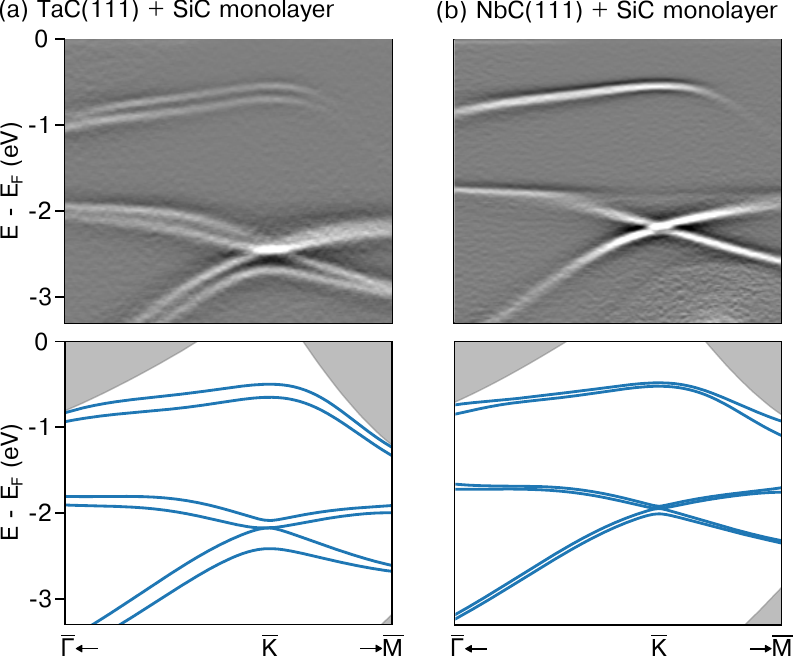}
	\caption{
		\footnotesize{(a) ARPES spectrum at h$\nu$=35~eV (T=80~K), presented as a second derivative image. Below is the calculated bandstructure. (b) The same for a SiC-terminated NbC film}
	}
	\label{fig:NbC}
\end{figure}

Close to the Dirac-like degeneracy at $\bar{K}$, it is possible to alternatively describe the dispersion using the k$\cdot$p model originally suggested for incorporating spin-orbit coupling effects into the graphene Dirac-cone dispersion\cite{Gmitra2009, KaneMele2005}:
\begin{equation}\label{eqn:F0}
\epsilon = \mu \lambda _{BR} + \nu \sqrt{(\hbar v k)^2 + (\lambda_{BR} - \mu \lambda_{I})^2}
\end{equation}
where $\nu$ = $\pm$1 indexes conduction and valence bands and $\mu$= $\pm$1 the spin-split bands, $v$ is the band curvature while 2$\lambda_{I}$ and 2$\lambda_{BR}$ are, respectively, the intrinsic spin-orbit splitting and extrinsic Bychkov-Rashba splitting. Good agreement with the experimental data is found using 2$\lambda_{I}$=60~meV and 2$\lambda_{BR}$=160~meV, which at $k=0$ (i.e., $\bar{K}$) corresponds to an absolute splitting in the valence bands of 220~meV. For reference, this is comparable to the largest reported band splittings observed in graphene following Au ($\approx$100~meV \cite{Marchenko2012,Marchenko2016}) or Pb intercalation ($\approx$200~meV \cite{Klimovskikh2017}). The sizable band splitting here derives from the heavy Ta atoms, readily confirmed by comparing with the same 2D-SiC termination that also forms on lighter niobium carbide films prepared in the same way (Figure \ref{fig:NbC})

The atomistic details of how this honeycomb SiC layer forms remain to be clarified. If we consider the SiC substrate in isolation, graphitization by thermal decomposition is a complex but by now well-understood phenomenon\cite{Shtepliuk2016}. At temperatures of $\approx$1400\textcelsius{} silicon atoms sublime from the surface while carbon, with a lower vapor pressure, does not. This leaves a carbon-rich layer that recrystallizes as graphene. In contrast, TaC is significantly more stable at high temperatures; to decompose it on a timescale of minutes would require extraordinary temperatures ($>$3000\textcelsius{}\cite{Kempter1960}) far from what has been reached here. Therefore we expect that the process is driven by the decomposition of the underlying SiC, leaving the TaC film intact. The most likely scenario is that the SiC termination forms during the post-deposition anneal of the TaC film to 1700\textcelsius{}, with Si and C diffusing through the TaC to accumulate on the surface. In this scenario, the sole function of the UHV annealing, at least for temperatures up to 1000\textcelsius{}, is to eliminate oxidation of the already-present SiC layer on the surface. This is consistent with our observation that the surface condition is unaffected by longer or additional UHV anneals. However, this does not hold for much higher UHV annealing temperatures ($>$1200\textcelsius{}), for which the surface termination transforms into graphene (see Supplementary Material). 

In conclusion, we have demonstrated a route to prepare large-area, high-quality SiC honeycomb monolayers through the annealing of thin films of transition metal carbides grown on 4H-SiC(0001). We have shown this for TaC and NbC, but anticipate that the process is generic and may also occur with other suitably lattice-matched rocksalt monocarbides on SiC such as VC, TiC, or WC. Band structure calculations provide good overall agreement with ARPES measurements. Together they reveal that the $\pi$ band of the SiC layer hybridizes with the underlying TaC surface states to yield a strongly spin-orbit split Dirac-like feature at the $\bar{K}$ points. Beyond seeking a clearer understanding of the formation mechanism, in the future it will be interesting to explore possibilities of modifying this termination by intercalation or adsorption of other species. Such experiments may decouple the SiC layer from the substrate, alter the band splitting or even bring the Dirac-like feature towards the Fermi energy. The bottom-up fabrication technique shown here finally brings 2D SiC monolayers into the growing family of honeycomb structured monolayers.

\begin{acknowledgments}
 We thank D. Carbone for assistance during photoemission measurements and useful discussions. This work was jointly supported by the Swedish Foundation for Strategic Research (SSF) (No. GMT14-0077, No.  RMA15-0024), Chalmers Excellence Initiative Nano, the Swedish Research council under Contract No. 2021-05252, 2D TECH VINNOVA competence Center (Ref. 2019-00068) and the Chalmers-MAX IV collaboration program. This work was performed in part at Myfab Chalmers. Research conducted at MAX IV, a Swedish national user facility, is supported by the Swedish Research council under Contract No. 2018-07152, the Swedish Governmental Agency for Innovation Systems under Contract No. 2018-04969, and Formas under Contract  No. 2019-02496.
\end{acknowledgments}

\bibliographystyle{unsrt}

\setlength{\voffset}{0cm}
\setlength{\hoffset}{0cm}

\newpage


\section*{Supplementary material}

# Supplementary Material: Bottom-up growth of monolayer honeycomb SiC

C. M. Polley,[1,2,*] H. Fedderwitz,[1] T. Balasubramanian,[1] A. A. Zakharov,[1] R. Yakimova,[3] J. Ekman,[2] O. Bäcke,[2] S. P. Dash,[2] S. Kubatkin,[2] and S. Lara-Avila[2]

[1]*MAX IV Laboratory, Lund University,*
*Fotongatan 2, 22484 Lund, Sweden*

[2]*Department of Microtechnology and Nanoscience,*
*Chalmers University of Technology, 412 96, Gothenburg, Sweden*

[3]*Department of Physics, Chemistry and Biology (IFM),*
*Linköping University, Linköping, 581 83 Sweden*

**METHODS**

**Sample preparation**. 4" wafers of high-purity semi-insulating 4H–SiC were obtained from Cree, from which (7×7) mm $^2$ substrates were cut. These were cleaned using a standard RCA wafer cleaning process. Immediately prior to sputter deposition, a substrate was cleaned in 2% HF for 3 minutes, rinsed in water, and blow-dried with nitrogen. Sputtering occured in a Nordiko 2000 magnetron sputtering system, pumped to a base pressure of 5×10$^{-7}$ mbar. The substrate was held at 350°C overnight before beginning. Sputter deposition of the carbide occurred at a power of 40 W, with an argon flow rate of 50 sccm and pressure of 5 mTorr, while the substrate temperature was approximately 800°C. This resulted in a deposition rate of ≈0.1 nm/s. After sputtering, the substrate was cooled at a rate of 30 °C/min. Following the sputter deposition, the film was annealed in two steps in a separate vacuum furnace apparatus with a base pressure of 5×10$^{-7}$ mbar. In the first stage the substrate temperature was increased from room temperature to 1150°C at a rate of 40 °C/min. At this point, the temperature was stabilized and argon introduced to a pressure of 800 mbar. Then the temperature was increased to 1700°C at a rate of 20 °C/min. This temperature was maintained for 5 minutes, after which the sample was cooled down to room temperature at 20 °C/min. For UHV studies, high-temperature annealing was accomplished by direct current heating, i.e. passing an electrical current through the sample itself. The sample temperature was estimated by an infrared pyrometer, and as such all temperatures stated should be considered only approximate to within 50-100°C.

**Band structure calculations** were performed within density-functional theory using the Perdew–Burke–Ernzerhof GGA exchange-correlation functional, as implemented in the Quantum ESPRESSO package[1,2]. Spin-orbit coupling was included through the use of fully relativistic projector augmented wave pseudopotentials. Band structures were calculated from a TaC slab model consisting of between 17 and 29 atomic TaC layers (15-30Å) plus terminating Si and C atoms, with 40Å of vacuum separation between slabs. The plane-wave cutoff energy and the k-point mesh sampling density were 70 Ry and 16×16×1, respectively. All discussed properties were checked for convergence with respect to slab thickness, vacuum cladding and k-point mesh. Lattice parameters were fixed to experimental values from literature, with the exception of slab models for which all atoms were fully relaxed along the c-axis.

2

**Photoemission measurements** were performed at the Bloch beamline at the MAX IV Laboratory.

**STM measurements** were performed at the Bloch beamline using a UHV coupled Omicron VT-XA. Images were acquired at room temperature with electrochemically etched tungsten tips.

**Electron microscopy**. Low energy electron microscopy measurements were performed at the MAXPEEM beamline at the MAX IV laboratory. Scanning transmission electron microscopy (STEM) was performed at Chalmers University of Technology. An FEI Versa 3D combined focused ion beam/scanning electron microscopy (FIB/SEM) workstation was used to prepare a thin foil cross section of the sample. In order to protect the sample surface from ion beam damage during sample preparation, a thin Pt layer was first deposited onto the surface using the electron beam and then a thicker Pt layer (2 $\mu$m) was deposited on top using Ga-ions. For most of the lift out procedure, an acceleration voltage of 30 kV was used for the ion beam. In order to reduce damage from the ion beam, the final thinning of the sample was done at lower acceleration voltages and currents, 5 kV and 50 pA followed by 2 kV and 20 pA. The thin foil sample was investigated by scanning transmission electron microscopy (STEM) using a FEI Titan 80–300 transmission electron microscope equipped with a Schottky FEG, operating at an accelerating voltage of 300 kV. STEM high resolution micrographs were acquired using a high angle annular dark-field (HAADF) and a bright-field (BF) detector. Micrographs were analyzed using the software TEM Imagining and Analysis.

**UHV SAMPLE PREPARATION**

The sample becomes oxidized when exposed to air, and a UHV annealing step is required to eliminate oxygen from the surface. In Figure 1 we demonstrate this evolution by following ARPES and core level spectra as a function of temperature. The initial anneal to 150°C serves only to desorb weakly physisorbed species but the surface remains covered with an amorphous oxide. There is no observable structure in LEED nor in ARPES (Fig 1a), and oxygen dominates the core level spectrum (Fig 1b, blue traces). Annealing to 870°C reduces one of the oxygen and carbon peaks (Fig 1b, orange traces), and at this point dispersive bands can be seen in ARPES. Further heating to 1000°C is sufficient to remove all oxygen from the surface. The Si 2p core level becomes much sharper and experiences a large (4eV) shift to lower binding energy, consistent with charge transfer due to the reduction. A second Ta 4f doublet appears, striking new surface states appear in ARPES and intense hexagonal spots are observed in LEED. This is the clean honeycomb SiC termination discussed at length in the main text.

In order to study the evolution with higher annealing temperatures, we can take advantage of a heat gradient produced by the UHV annealing process. This gradient results from localized hot-spots at two opposite corners of the sample where it is mechanically clamped, due to contact and spreading resistance. While it is difficult to accurately measure the temperature in these corners with an optical pyrometer, it is clear by eye that the corners are several hundred degrees hotter than the ≈1000°C center of the sample.

The evolution in this regime is most clearly understood by examining ARPES spectra (Fig 1c). As the annealing temperature is increased the spin-split surface states are eliminated and graphene $\pi$ bands appear - first monolayer then bilayer (note that the graphene unit cell is rotated by 30° with respect to the TaC and SiC, as is also the case for regular graphene growth on SiC). The bulk bandstructure of the TaC film remains visible in all cases. This evolution is also reflected in LEED (Fig 1f), with the emergence of new spots corresponding to graphene when probing the sample corners. The broadness of these new spots indicates minor azimuthal disorder of the graphene layers. Core level spectra (Fig 1d) demonstrate that when moving towards hotter parts of the sample the Si 2p doublet and C 1s component $HC$ from the honeycomb SiC layer are eliminated, along with the surface Ta 4f doublet. Meanwhile the C 1s spectra acquire new features $G1$ and $G2$ corresponding to monolayer

4

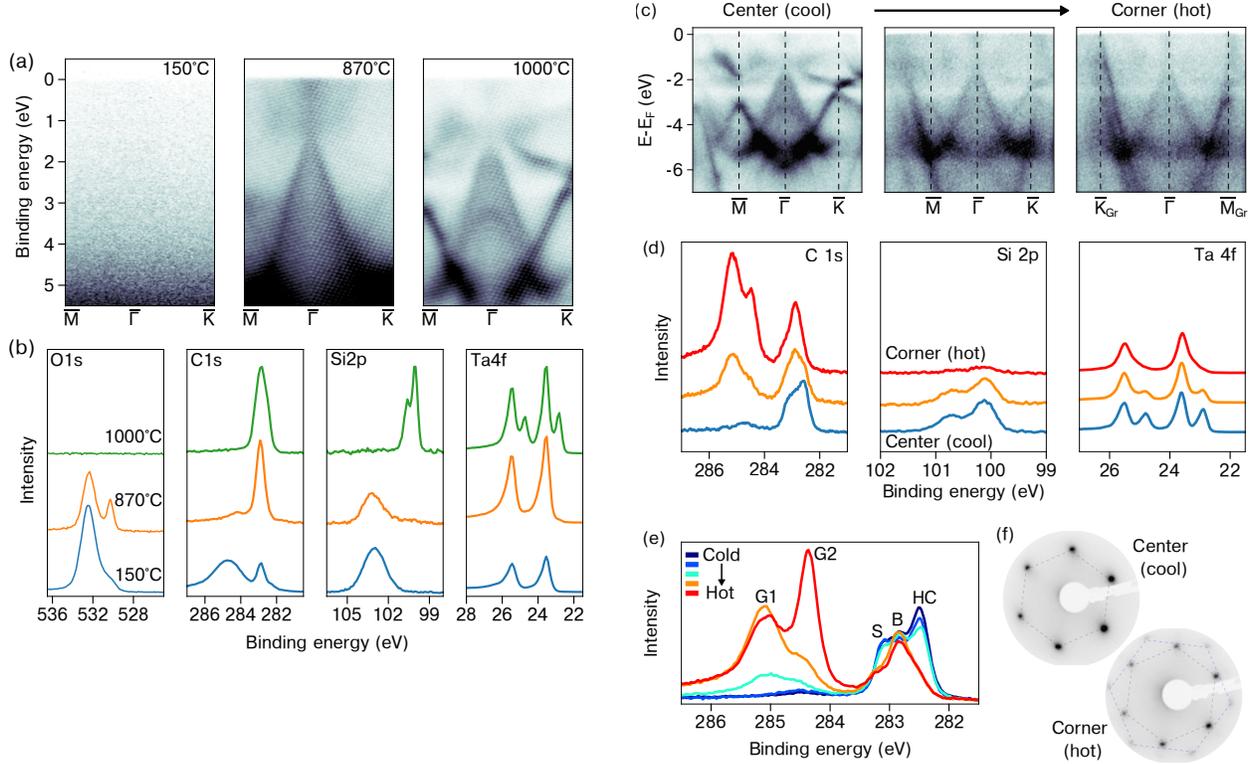

FIG. 1. **Evolution of the surface termination with increasing annealing temperature** (a) ARPES spectra (hν=180 eV, T=80 K) and (b) corresponding core level spectroscopy (hν=650 eV) as the sample is annealed to progressively higher temperatures in UHV. Even higher temperature scales are accessed by moving from the center towards one of the two mounting corners of the sample. (c) High symmetry ARPES cuts through the surface Brillouin zone (T=50 K, hν=180 eV). (d) Corresponding C 1s, Si 2p and Ta 4f core levels (T=300 K, hν=430 eV). (e) Higher resolution C 1s spectra (T=50 K, hν=430 eV). (f) LEED images, showing the emergence of graphene spots when moving to hotter regions of the sample

and bilayer graphene. The modifications to the C1s spectrum are more clearly seen in the higher resolution spectra in Figure 1e. The measurements in Figures 1d and 1e are taken from separate experimental runs and hence do not match exactly.

Note that it is the non-uniformity of the UHV annealing method that creates this variation across the surface; the as-grown films are uniform.

**SURFACE STATES OF CLEAN TAC(111)**

Figure 2 presents DFT slab bandstructure calculations for a clean TaC film, with a C-atom termination (a) and the preferred Ta termination (b). For the Ta termination, two spin-split states are found within the projected bandgap at $\bar{K}$, that are good agreement with an earlier ARPES study by Anazawa[3]. Based on this we can exclude the possibility that the bandstructure discussed in the main text simply derives from a clean TaC(111) surface.

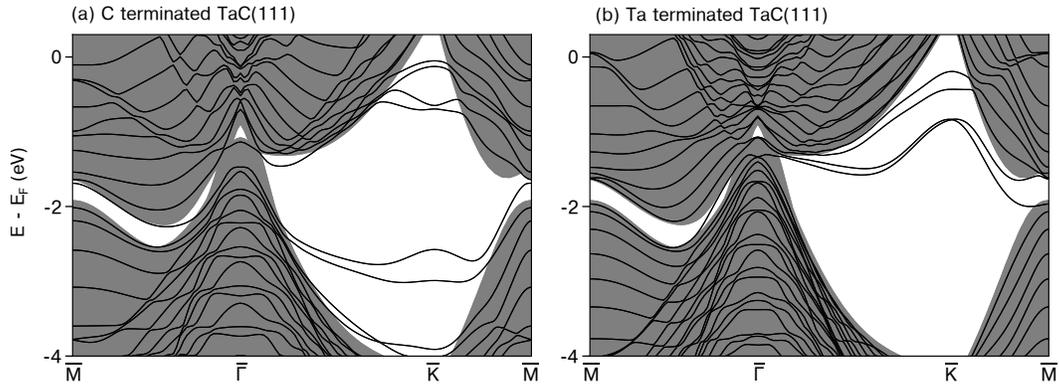

FIG. 2. **Surface states on clean TaC(111)** (a) DFT bandstructure calculation for a 23-layer C terminated TaC slab (b) The same for a 31-layer Ta terminated slab

**SLAB THICKNESS AND SPIN-ORBIT COUPLING**

In Figure 3 we compare DFT slab bandstructure calculations for SiC terminated TaC(111), with the structure described in the main text, for various slab thicknesses and with spin orbit coupling excluded (a) or included (b). In these plots we have not reduced the buckling in the SiC layer from the relaxed value, hence the bands remain gapped at $\bar{K}$. While the bandstructure near $\bar{M}$ requires very thick slabs to converge, around $\bar{K}$ it converges quickly.

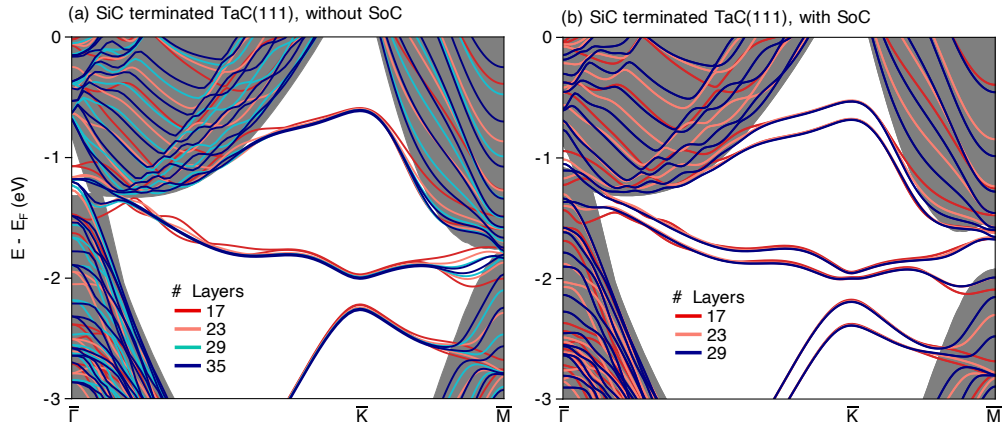

FIG. 3. **Thickness dependent DFT slab calculations**. Shown with spin-orbit coupling (a) excluded and (b) included

## BUCKLING OF THE SIC MONOLAYER

As discussed in the main text, in order to reproduce the degeneracy observed experimentally at $\bar{K}$, the buckling of the SiC monolayer must be reduced by 5pm compared to the DFT relaxed structure. This is summarized in Figure 4.

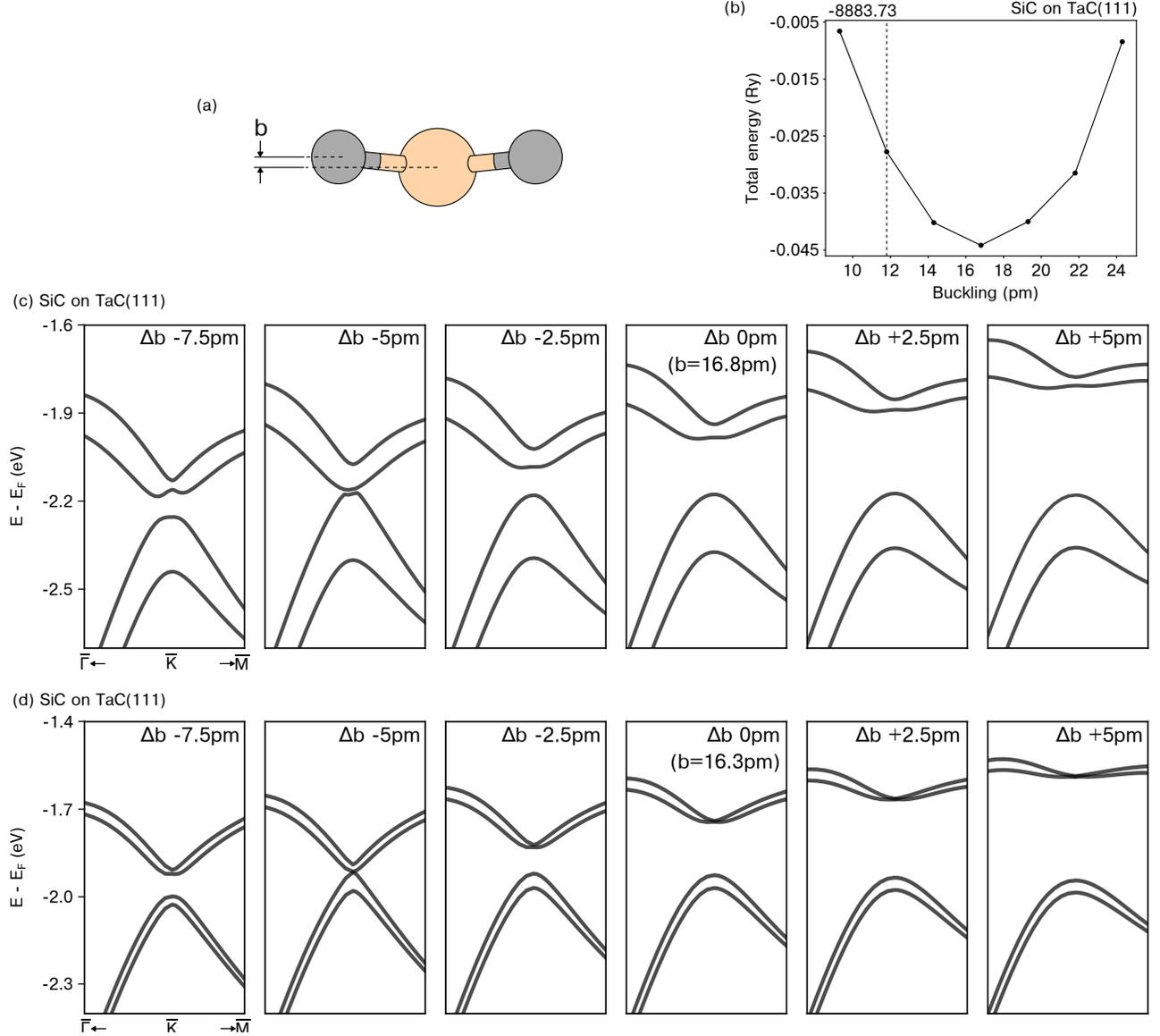

FIG. 4. **Role of buckling in the DFT-predicted bandgap**. By manually adjusting the buckling of the SiC monolayer (defined as parameter $b$ in (a)) away from the relaxed, lowest total energy state (b), it is possible to close the bandgap at $\bar{K}$ to better match experimental observations. This is true for SiC monolayers on both (c) TaC and (d) NbC. Here $\Delta b$ represents the deviation in buckling from the value predicted by relaxation calculations.

8

Experimentally, all of these values are still far larger than what is observed in line profiles from atomic resolution STM topography maps, shown in Figure 5. We note however that interpreting such measurements requires caution since the apparent buckling can change depending on the imaging conditions.

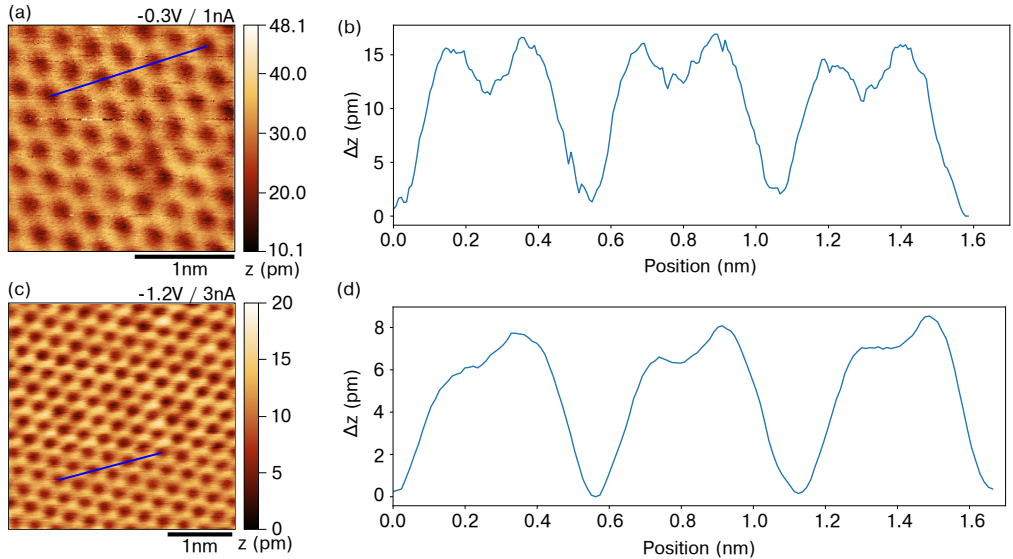

FIG. 5. **STM height profiles**. Line profiles through atomic resolution STM topography maps (profile indicated by blue line on (a) and (c)) indicate buckling levels well below 5pm

## MULTI-SCALE TOPOGRAPHY MAPS

The ARPES measurements presented in the main text are spatially averaged over the photon beam footprint on the sample of (10 $\mu$m × 15 $\mu$m). No variation in the spectra could be observed when scanning the sample across the beam on millimeter scales. To investigate the surface with higher spatial resolution, Figure 6 presents microscopy maps of the same TaC(111) film surface for length scales spanning 10 $\mu$m through 2 nm, using low energy electron microscopy (LEEM), atomic force microscopy (AFM) and scanning tunneling microscopy (STM). The surface is characterized by uniform and atomically flat terraces, that are divided by cracks into long parallel strips approximately 200 nm wide. These cracks are approximately 5-20 nm wide and 1 nm deep, and are believed to correspond to atomic steps from the underlying vicinal SiC substrate. Owing to a phase contrast effect, these cracks appear as bright stripes in the otherwise uniform LEEM image. The LEEM image was

acquired at the MAXPEEM endstation at the MAX IV Laboratory, after a 1050°C anneal in UHV conditions. All STM images were acquired on the UHV-coupled STM of the Bloch ARPES beamline at the MAX IV Laboratory.

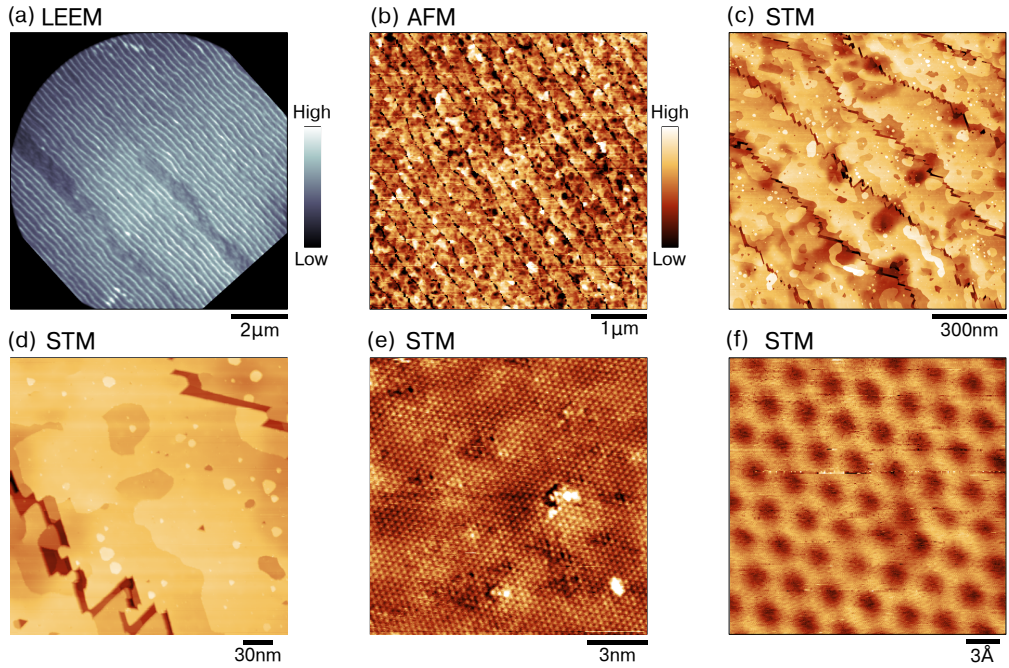

FIG. 6. **Topography maps of the TaC(111) surface** (a) LEEM image (10 $\mu$m field of view), acquired in UHV after annealing to 1050°C. (b) AFM image (5 $\mu$m frame) of the oxidized surface. (c-f) Progressively higher magnification STM images of the clean surface, (c,d,e: $V_{bias}$ = -1.5 V / I = 1.5 nA, f: $V_{bias}$ = -0.3 V / I = 1 nA)

10

**SUPERCONDUCTIVITY**

Figure 7 presents a temperature dependent two-terminal resistance curve for a representative TaC(111) film, measured outside of UHV. The observed superconducting transition temperature of 9.1 K is consistent with stoichiometric, cubic TaC[4].

Electrical characterization was performed in a helium cryostat capable of temperatures down to 2 K and magnetic fields up to 14 T. Measurements were performed with a constant DC current of 100 nA using a Keithley 6221 source and Agilent 34420A nanovolt meter.

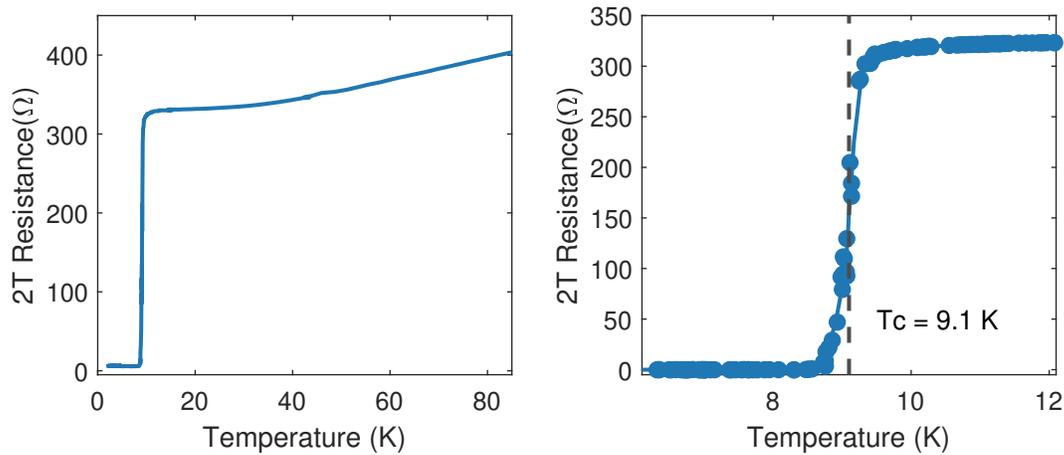

FIG. 7. **Superconductivity of TaC(111) film** (a) Temperature dependent resistance of a TaC(111) film (I=100 nA) (b) The same data, magnified around the transition temperature

## STRUCTURAL TRANSFORMATION DURING POST-DEPOSITION ANNEAL

X-ray diffraction (XRD) spectra from a representative TaC(111) film (Figure 8), acquired with a CuK$\alpha$ X-ray source ($\lambda$=1.5418 Å), demonstrate the role of the 1700℃ post-deposition anneal in establishing structural ordering of the TaC film. Prior to this anneal diffraction peaks are only observed from the 4H-SiC substrate. After annealing, new (111) and (222) peaks originating from crystalline TaC emerge. No superconductivity is observed in the films if this post-annealing step is not performed.

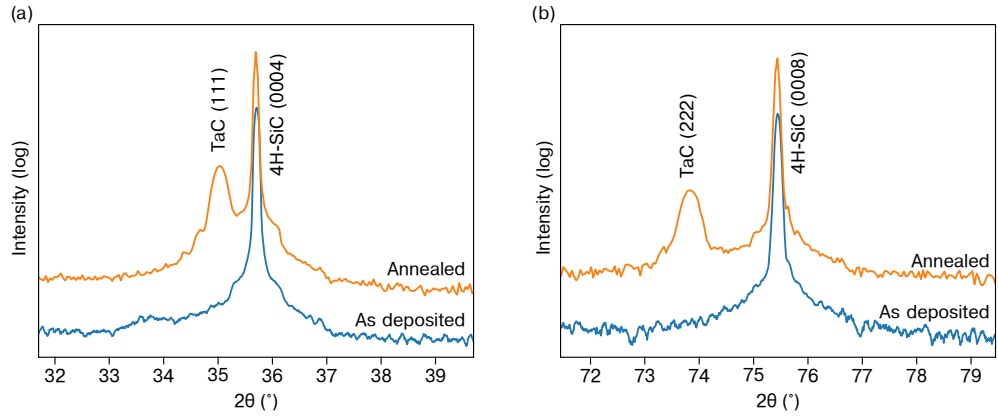

FIG. 8. **XRD of TaC(111) film** XRD spectra indicating the structural ordering of the TaC film following the post-deposition anneal to 1700℃.

## ALTERNATIVE STRUCTURES AND REGISTRIES

Here we summarize bandstructure calculations for different registries and different structures. The structure assumed in the main text (Fig. 9c) is highlighted, and is considered to be correct on the basis of providing the best agreement with the experimental measurements. All calculations are overlaid on a second derivative image of a spectrum acquired at 36eV.

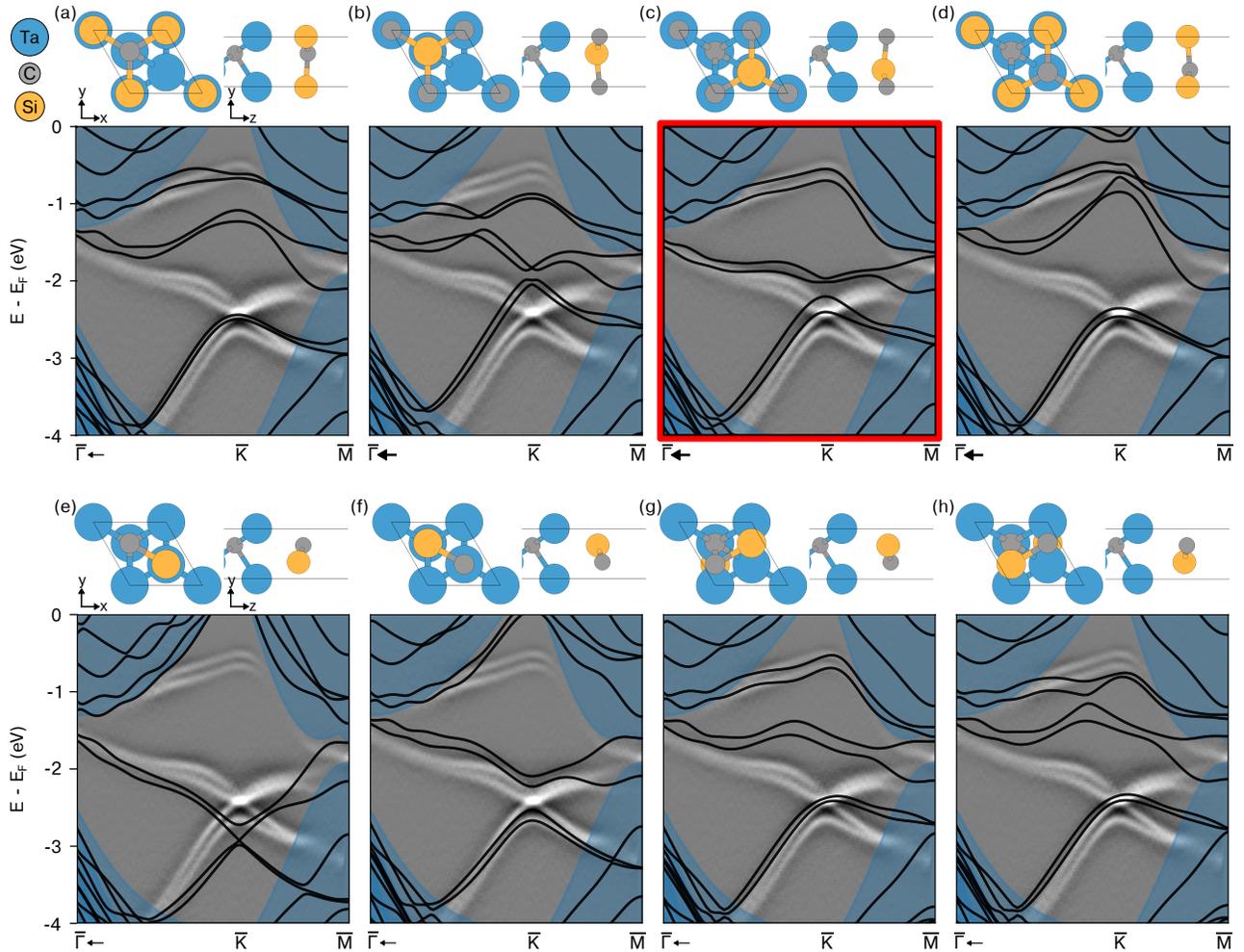

FIG. 9. **Bandstructure calculations for a honeycomb silicon carbide monolayer on a Ta-terminated TaC(111) slab**. Panels (a-h) simulate different registries of the SiC layer. The highlighted registry (c) has been assumed in the main text.

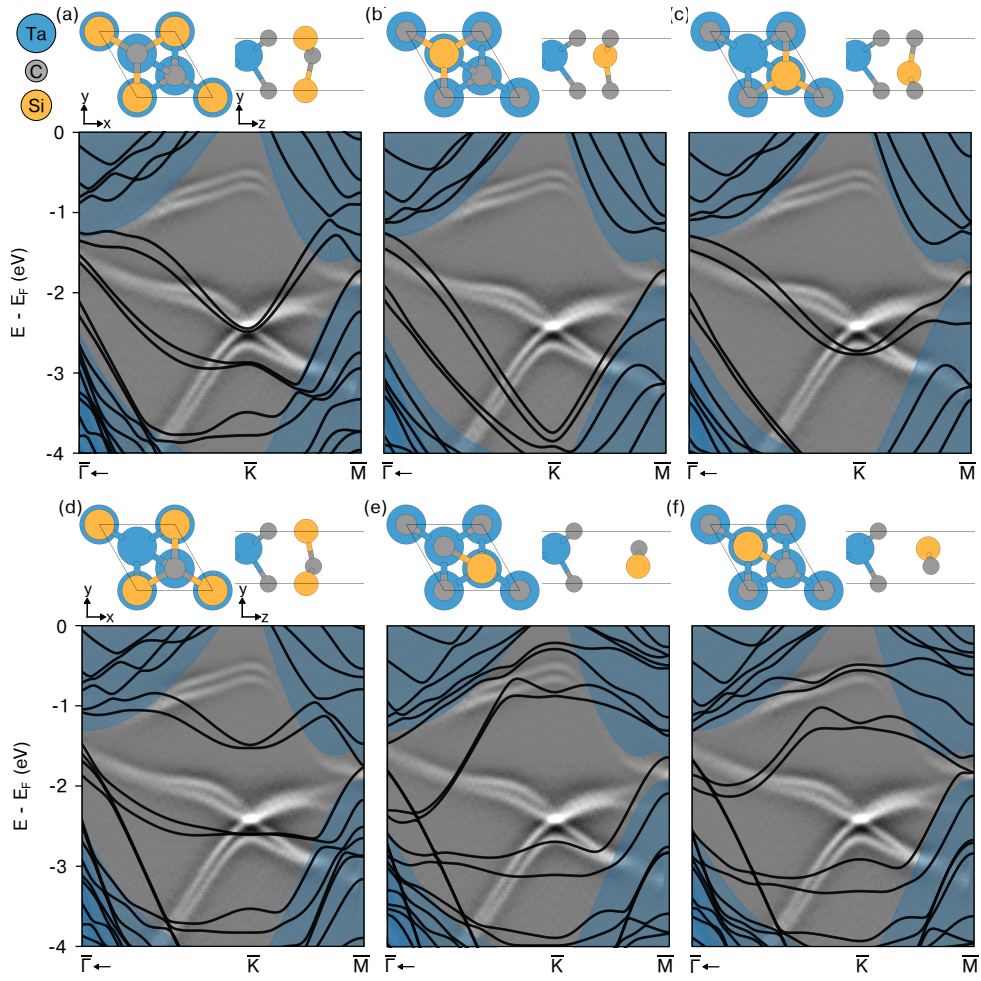

FIG. 10. **Bandstructure calculations for a honeycomb silicon carbide monolayer on a C-terminated TaC(111) slab**. Panels (a-f) simulate different registries of the SiC layer. Note that carbon is not the preferred termination for TaC(111).

14

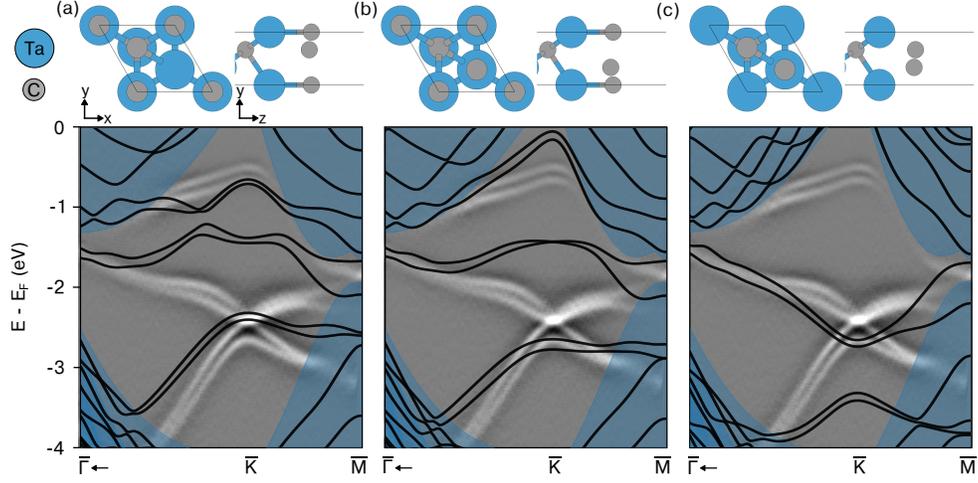

FIG. 11. **Bandstructure calculations for a honeycomb carbon monolayer on a Ta-terminated TaC(111) slab**. Panels (a-c) simulate different registries of the carbon layer.

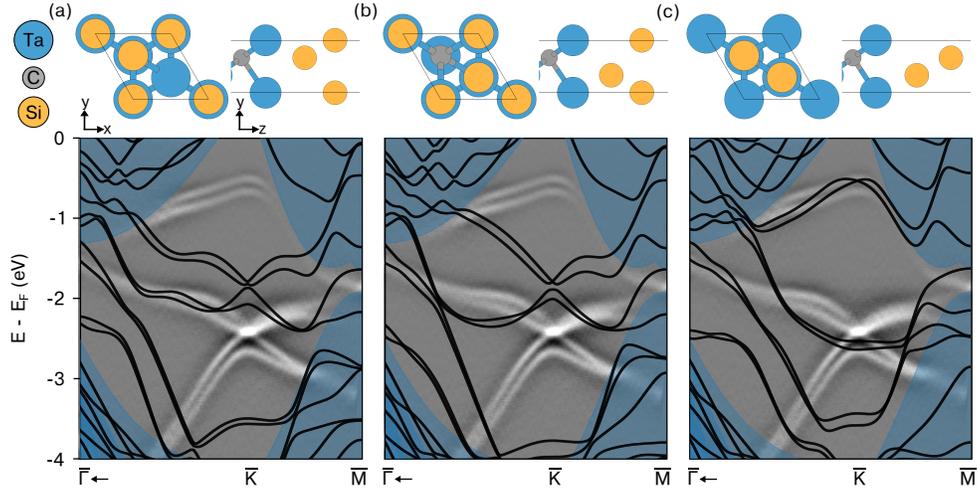

FIG. 12. **Bandstructure calculations for a honeycomb silicon monolayer on a Ta-terminated TaC(111) slab**. Panels (a-c) simulate different registries of the silicon layer.

15

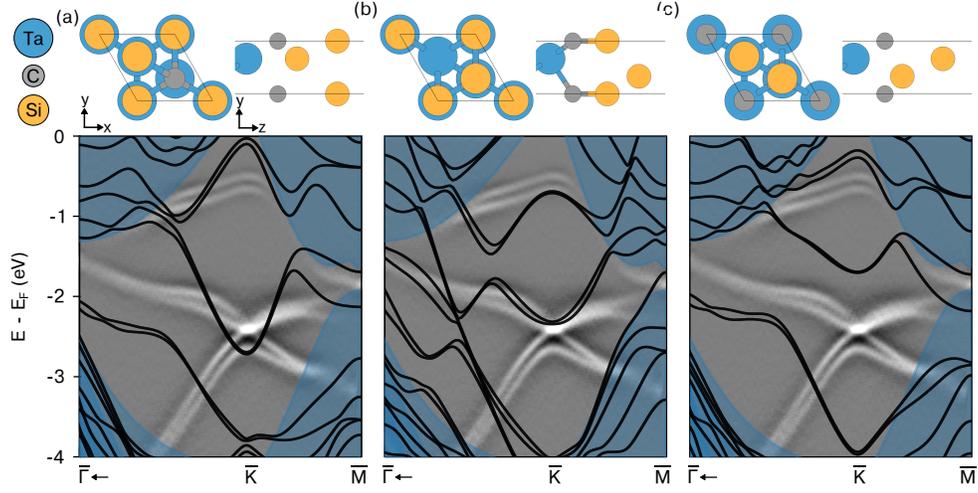

FIG. 13. **Bandstructure calculations for a honeycomb silicon monolayer on a C-terminated TaC(111) slab**. Panels (a-c) simulate different registries of the silicon layer.

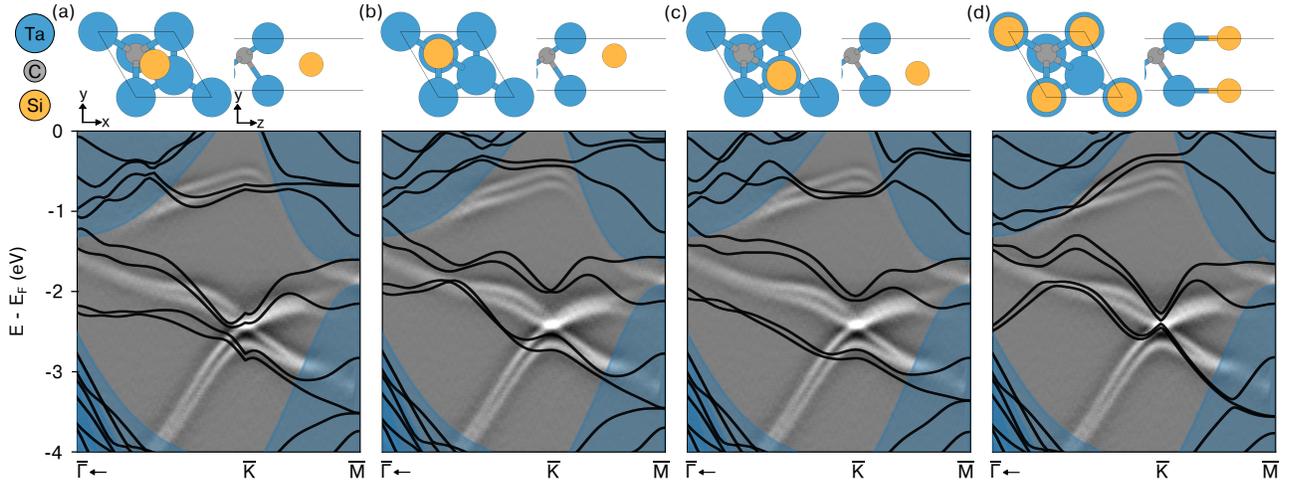

FIG. 14. **Bandstructure calculations for a single silicon adatom atop a Ta-terminated TaC(111) slab**. Panels (a-d) simulate different sites for the silicon atom.

16



\end{document}